\documentclass[aps,prl,twocolumn,superscriptaddress,groupedaddress]{revtex4}
\usepackage{epstopdf}
\pdfoutput=1
\usepackage[utf8]{inputenc}
\usepackage[T1]{fontenc}
\usepackage{lmodern,pgf}
\usepackage[british]{babel}
\usepackage{amsmath}
\usepackage{amssymb}
\usepackage{amsfonts}
\usepackage{ulem,bbm}
\usepackage{color}
\usepackage{float}
\usepackage{appendix}
\usepackage{graphicx}
\usepackage{grffile}
\usepackage{hyperref}
\usepackage{textcomp}
\usepackage{subfigure}
\usepackage{booktabs}

\newcommand*\samethanks[1][\value{footnote}]{\footnotemark[#1]}

\newcommand{\I}{\ensuremath{\mathrm{i}\hspace{1pt}}}
\newcommand{\api}{\ensuremath{\text{a-}\pi}}
\newcommand{\aetap}{\ensuremath{\text{a-}\eta'}}
\newcommand{\afn}{\ensuremath{\text{a-}f_0}}

\renewcommand{\lambdabar}{\bar{\lambda}}
\newcommand{\wchiral}{w_{0,\chi}}
\newcommand{\arxiv}[2]{[arXiv:\,\href{http://arxiv.org/abs/#1}{\texttt{#1}} [\texttt{#2}]]}
\newcommand{\arxivold}[1]{[arXiv:\,\href{http://arxiv.org/abs/#1}{\texttt{#1}}\,]}

\begin{document}
\textmd{MS-TP-19-04}
\title{Numerical results for the lightest bound states in $\boldmath{\mathcal{N}=1}$ 
supersymmetric SU(3) Yang-Mills theory}
\author{Sajid Ali%
\thanks{\{sajid.ali,h.gerber,munsteg\}@uni-muenster.de}}
\affiliation{University of M\"unster, Institute for Theoretical Physics, 
Wilhelm-Klemm-Str.~9, D-48149 M\"unster, Germany}
\affiliation{Government College University Lahore, Department of Physics,
Lahore 54000, Pakistan}
\author{Georg Bergner\thanks{georg.bergner@uni-jena.de}}
\affiliation{University of Jena, Institute for Theoretical Physics, 
Max-Wien-Platz 1, D-07743 Jena, Germany}
\address{University of M\"unster, Institute for Theoretical Physics, 
Wilhelm-Klemm-Str.~9, D-48149 M\"unster, Germany}
\author{Henning Gerber\samethanks[1]}
\affiliation{University of M\"unster, Institute for Theoretical Physics, 
Wilhelm-Klemm-Str.~9, D-48149 M\"unster, Germany}
\author{Istvan Montvay\thanks{montvay@mail.desy.de}}
\affiliation{Deutsches Elektronen-Synchrotron DESY, Notkestr.~85, D-22607 Hamburg, 
Germany}
\author{Gernot M\"unster\samethanks[1]}
\affiliation{University of M\"unster, Institute for Theoretical Physics, 
Wilhelm-Klemm-Str.~9, D-48149 M\"unster, Germany}
\author{Stefano Piemonte\thanks{stefano.piemonte@ur.de}}
\affiliation{University of Regensburg, Institute for Theoretical Physics, 
Universit\"atsstr.~31, D-93040 Regensburg, Germany}
\author{Philipp Scior\thanks{scior@physik.uni-bielefeld.de}}
\affiliation{University of Bielefeld, Faculty of Physics,
Universit\"atsstr.~25, D-33615 Bielefeld, Germany}
                      
\date{\today}

\begin{abstract}
The physical particles in supersymmetric Yang-Mills theory (SYM) are bound states of gluons and gluinos. We have determined the masses of the lightest bound states in  SU(3) $\mathcal{N}=1$ SYM. Our simulations cover a range of different lattice spacings, which for the first time allows an extrapolation to the continuum limit. Our results show the formation of a supermultiplet of bound states, which provides a clear evidence for unbroken supersymmetry.
\end{abstract}

\maketitle


Supersymmetry (SUSY) plays a fundamental role in the physics of elementary particles beyond the Standard Model. The understanding of the non-perturbative phenomena of SUSY theories is important since they might explain the supersymmetry breaking at low energies. Besides the relevance for extensions of the Standard Model, supersymmetric gauge theories also provide insights into non-perturbative phenomena that also occur in QCD, such as confinement of color charges, at least in certain regimes since supersymmetry constrains the non-perturbative contributions. Non-perturbative numerical methods such as lattice simulations are essential to complement and extend the obtained analytical understanding from SUSY models to theories with less or no supersymmetry.

Supersymmetric extensions of the Standard Model must include the superpartners of the gluons, the so-called gluinos, which are Majorana fermions transforming under the adjoint (octet) representation of SU(3). The gluino would interact strongly, and the minimal theory describing the interactions between gluons and gluinos is $\mathcal{N}=1$ supersymmetric SU(3) Yang-Mills theory, abbreviated SU(3) SYM.
The strong interactions between gluons and gluinos are expected to give rise to bound states organised in supermultiplets degenerate in their masses, if supersymmetry is unbroken. The structure of the supermultiplets has been theoretically investigated in~\cite{Veneziano:1982ah,Farrar:1997fn,Farrar:1998rm}. The boson-fermion degeneracy is expected to appear at the non-perturbative level and, as a consequence, the singlet mesons and glueballs of QCD-like theories have an exotic fermion superpartner, the gluino-glue, which is a bound state of a single valence gluino with gluons.

SU(3) SYM is of a complexity comparable to QCD, and Monte Carlo lattice simulations are an ideal ab initio approach to investigate this theory. In particular, a study of the mass gap of the particle spectrum requires numerical simulations. As supersymmetry is explicitly broken by any lattice discretisation~\cite{Dondi:1976tx,Kato:2008sp,Bergner:2009vg,Bergner:2016sbv},
it is a challenging task to show that the bound states masses are consistent with the formation of supermultiplets in the continuum limit. It would open up the possibility of much further reaching numerical investigations of SYM and correspond to the first step towards a numerical investigation of supersymmetic QCD and gauge theories with extended supersymmetry, since SYM is one sector of these theories. Such a result would also provide evidence for the correctness of the conjectured bound state spectrum and for the absence of an unexpected breaking of supersymmetry by the non-perturbative dynamics. 

In this contribution we focus on the spectrum of bound states of the $\mathcal{N}=1$ supersymmetric Yang-Mills theory with gauge group SU(3). In previous projects we have investigated SYM with gauge group SU(2)~\cite{Bergner:2013nwa,Bergner:2015adz,Ali:2019gzj}, which can
be considered to be a test case for the more realistic SU(3) SYM that
contains the gluons of QCD. The gauge group SU(3) brings new physical
aspects; for instance, it has complex representations in contrast
to SU(2), and other types of bound states are possible. The breaking pattern
of the global chiral symmetry group is also quite different from the case of
SU(2). In particular, in the region of spontaneously broken symmetry it is
expected that CP-violating phases exist, which are related to each other by
discrete $Z_3$ transformations.

We have presented our first data at a single lattice spacing in~\cite{Ali:2018dnd} together 
with some estimates of systematic uncertainties.
The present work is the first final analysis for the lowest chiral supermultiplets of SU(3) SYM with 
a complete chiral and continuum extrapolation.


In the continuum the (on-shell) Lagrangian of SU(3) supersymmetric
Yang-Mills theory, containing the gluon fields $A_{\mu}$ and the gluino
field $\lambda$, is
\begin{equation}
\mathcal{L} = \mathrm{tr} \left[-\frac{1}{2}
F_{\mu\nu} F^{\mu\nu} + \I \bar{\lambda} \gamma^\mu
D_\mu \lambda -m_0 \bar{\lambda} \lambda \right] \,,
\end{equation}
where $F_{\mu\nu}$ is the non-Abelian field strength and $D_\mu$ denotes the
gauge covariant derivative in the adjoint representation of SU(3). The gluino
mass term with the bare mass parameter $m_0$ breaks supersymmetry softly.
The gauge coupling $g$ is represented in terms of $\beta=\frac{6}{g^2}$, and the mass in terms of the hopping parameter $\kappa=\frac{1}{2(m_0+4)}$.

The technical details of our approach for the numerical simulations of SU(3)
SYM have been described in our previous
publication~\cite{Ali:2018dnd}. We employ the lattice discretization of SYM
proposed by Curci and Veneziano \cite{Curci:1986sm}. In our approach the bare mass parameter 
is tuned to the chiral limit determined by the point where the adjoint pion $m_{\api}$ mass vanishes.
The basic Wilson action for the gluino is in our case improved by the clover term to
reduce the leading order lattice artefacts, see~\cite{Ali:2018dnd} for
further details. We have used the one-loop value for the coefficient
$c_{sw}$~\cite{Musberg:2013foa}, leading to a one-loop $O(a)$ improved
lattice action at finite lattice spacings $a$. 
As indicated by our first results~\cite{Ali:2018dnd}, the perturbative $c_{sw}$ is 
already sufficient to provide a drastic reduction of lattice artefacts even at quite coarse lattice spacings.

Alternative approaches have been investigated for the simulation of SYM~\cite{Giedt:2008xm,Endres:2009yp,Kim:2011fw,Steinhauser:2018wep}, but so far 
they did not succeed in the continuum extrapolation of the bound state spectrum.

The complexity and the cost of the numerical lattice
simulations for this theory is at least as challenging as in corresponding
investigations of QCD. Additionally, there are more specific challenges for
the technical realisation of numerical simulations of SYM, such as
the unavoidable explicit breaking of supersymmetry on the lattice. Therefore, the most important task of our project is to
demonstrate that the infrared physics emerging from the numerical simulations is consistent with
restoration of supersymmetry in the continuum limit.

A further specific challenge is related to the integration of Majorana fermions, 
which leads to an additional sign factor in the simulation~\cite{Ali:2018dnd}. This Pfaffian 
sign has to be considered in a reweighting of the observables.


\begin{figure*}
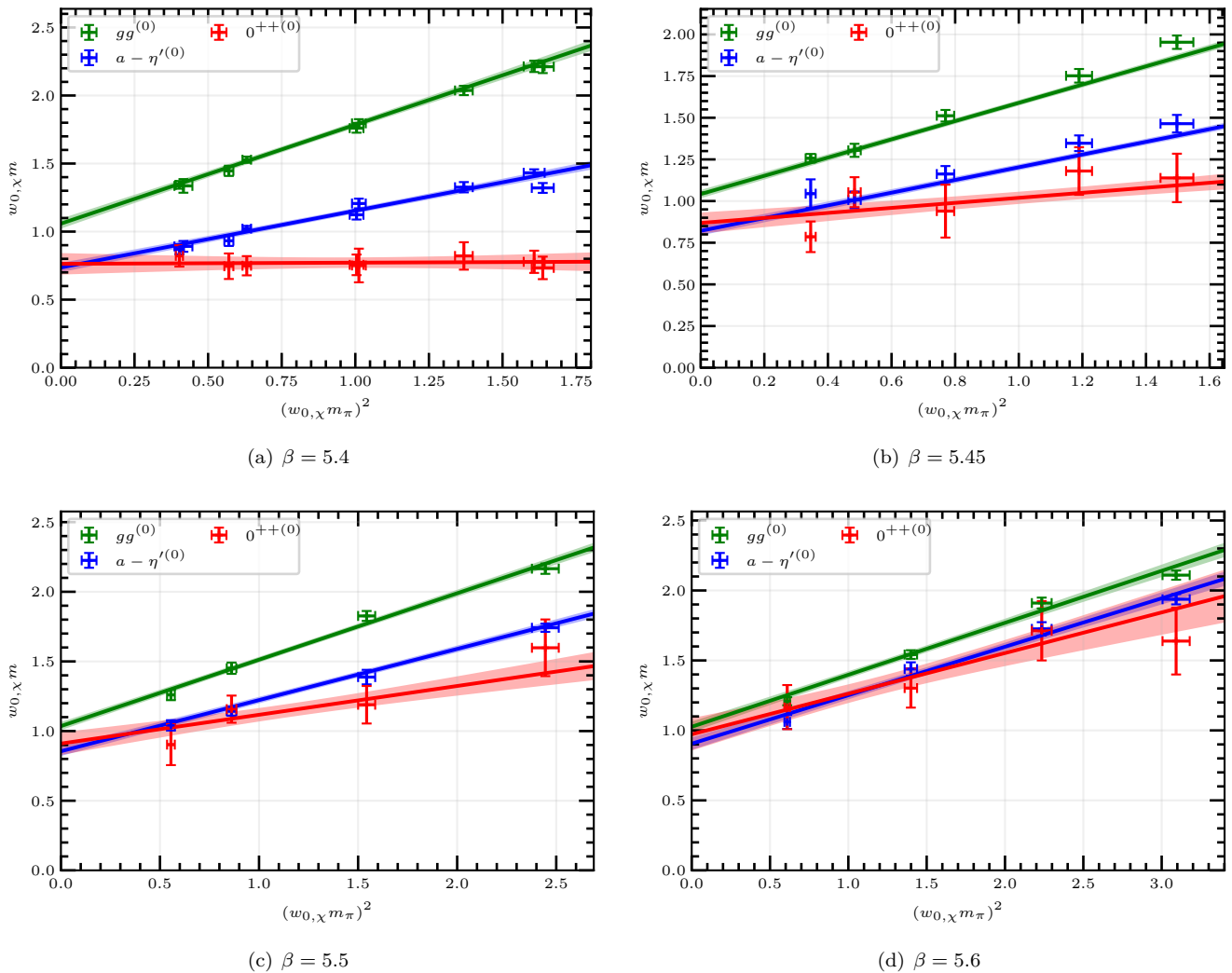

\subfigure[$\beta=5.4$]{\input{figures/5.4/chiralplot.pgf}}
\subfigure[$\beta=5.45$]{\input{figures/5.45/chiralplot.pgf}}
\subfigure[$\beta=5.5$]{\input{figures/5.5/chiralplot.pgf}}
\subfigure[$\beta=5.6$]{\input{figures/5.6/chiralplot.pgf}}
\caption{The chiral extrapolations of the particle masses at the different
 lattice spacings using the fit function \eqref{eq:linearFit} ($y=\frac{a}{\beta^2}$). The gluino-glue, the pseudoscalar $\aetap$ meson,
and the scalar channel, which includes a mixing of the glueball and the
$\afn$ meson, are extrapolated to the point where the adjoint pion mass
vanishes. }
\label{fig:chiralcoarse}
\end{figure*}

\begin{figure*}
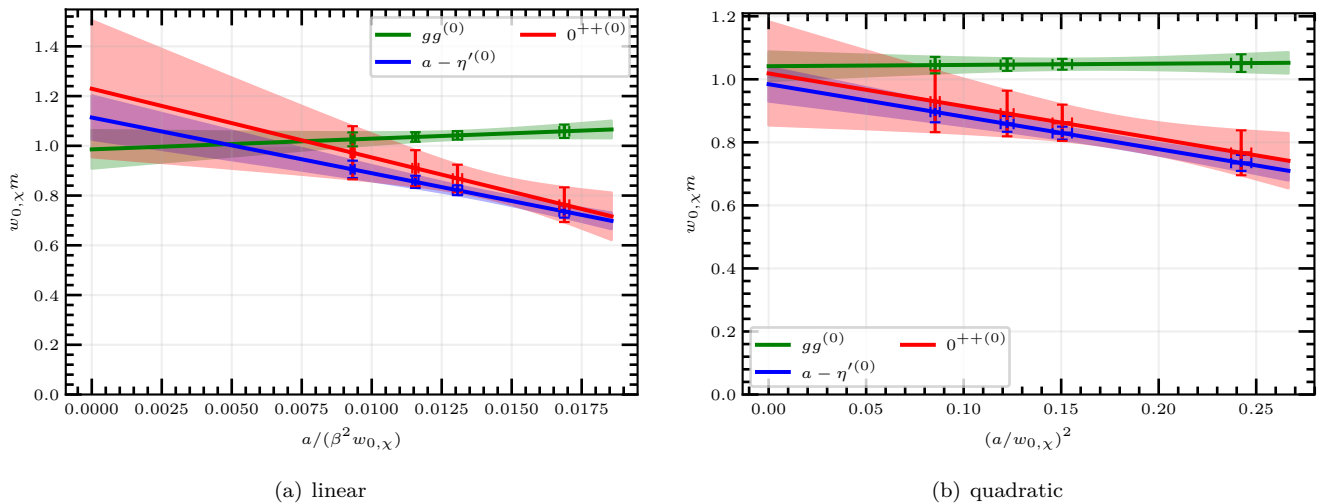

\hspace*{-1.4cm}
\begingroup
\makeatletter
\renewcommand{\p@subfigure}{}
\makeatother
\subfigure[linear\label{fig:continuumextr:a}]{\input{figures/continuum/linear.pgf}}
\subfigure[quadratic\label{fig:continuumextr:b}]{\input{figures/continuum/quadr.pgf}}
\caption{The extrapolation of the bound states masses to the continuum using the fit function \eqref{eq:linearFit} in \ref{fig:continuumextr:a} linear ($y=\frac{a}{\beta^2}$) and in \ref{fig:continuumextr:b} quadratic ($y=a^2$) in the lattice spacing. The datapoints mark the chirally extrapolated values at simulated lattice spacings. Since the chiral and continuum extrapolation is done simultaneously, they align completely with the fit curves.} 
\label{fig:continuumextr}
\endgroup
\end{figure*}
The scale, i.\,e.\ the determination of the lattice spacings in physical units in terms of a common observable, is measured from gluonic observables. We are using chirally extrapolated values of the scale $w_0$ from the gradient flow~\cite{Luscher:2010iy,Borsanyi:2012zs}\footnote{We have chosen the common reference value of $u=0.3$ ($w_0^{0.3}$) as explained
in~\cite{Bergner:2014ska}. Due to the insufficient topological sampling, a
value of $u=0.2$ has instead been chosen in our previous
work~\cite{Ali:2018dnd}. In our current work, the effect of the insufficient
topological sampling is under better control and there is no need for
the smaller value of $u$.}.  The chiral values $\wchiral$ are obtained at each $\beta$ by a fit of the data to a 
second order polynomial in the square of the adjoint pion mass in lattice units $(a m_{\api})^2$.

An improvement with respect to our work on SU(2) SYM, where we extrapolated the observables first to the chiral limit and in a second step 
to the continuum limit, is that we now use a combined fit towards the chiral and continuum limit. The  chiral continuum values $O_{\chi,\mathrm{cont.}}$ of the 
observable $O$ in units of $\wchiral$ are determined by
\begin{equation}\label{eq:linearFit}
O(m_{\api}^2,\wchiral)=O_{\chi,\mathrm{cont.}}+c^{(1)}x+c^{(2)}y+c^{(3)}xy,
\end{equation}
where $x=(\wchiral m_{\api})^2$ and $y=\frac{ a }{\wchiral \beta^2}$ (linear extrapolation). Due to the one-loop clover improvement of the action, we expect leading lattice artifacts to be of $\mathcal{O}(a/\beta^2)$ for on-shell observables, which leads to the dependence on the gauge coupling in $y$.
The $\mathcal{O}(a/\beta^2)$ contribution could, however, be very small since considerable improvements have been observed already with the tuning to the one-loop level.  In order to compare both cases, we perform additional fits with the leading lattice artifact term $\mathcal{O}(a^2)$, i.\,e.\ $y=\frac{a^2}{\wchiral^2}$ in~\eqref{eq:linearFit} (quadratic extrapolation).

The main indication for restoration of supersymmetry in lattice
simulations presented in this paper is the formation of mass degenerate
supermultiplets. An alternative indication is given by the supersymmetric
Ward identities. The violation of the supersymmetric Ward identities in the
chiral limit is an indication of lattice artefacts, since chiral symmetry
and supersymmetry should be restored at the same point in the continuum
theory, if there is no unexpected supersymmetry breaking. 
The Ward identities also provide a cross check for the tuning of the bare gluino mass parameter.
We have found that the Ward identities are consistent with a restoration of supersymmetry, and the leading lattice artefacts are $\mathcal{O}(a^2)$ as found in~\cite{Ali:2018fbq}. This analysis will soon appear in a separate publication.

We have performed simulations at a large range of values of the inverse gauge coupling $\beta$ ranging from $\beta=5.2$ up to $\beta=5.8$ to search for an optimal window for the continuum limit extrapolation. 
In our previous work~\cite{Ali:2018dnd} we have presented the first results
for the particle spectrum of SU(3) SYM obtained at a single lattice spacing. We have now investigated the
systematic uncertainties regarding the finite size effects, the sampling of
topological sectors, and the fluctuations of the Pfaffian sign, and found a
parameter range where these effects are under control. Only a subset of the considered $\beta$ range turned out to be reliable for the determination of the
bound states. The coarsest lattice spacings (smallest $\beta$ values) are too far away from the continuum limit, which makes the extrapolation unreliable. The finest lattice spacings (largest $\beta$ values) suffer from large finite volume effects and a freezing of the topological fluctuations. According to these criteria, our final selection of $\beta$ values is $5.4$, $5.45$, $5.5$, and $5.6$. 

In the current work we present the final results for the lightest particles
of SU(3) SYM. We are now able to combine several different lattice spacings
in an extrapolation to the continuum limit. 
In comparison to~\cite{Ali:2018dnd}, we have also improved our determination of the bound states,
leading to a clearer signal for the particle masses.
These methods have been introduced and tested with the data of SU(2) SYM in~\cite{Ali:2019gzj}.

The considered states and corresponding interpolating operators are the
scalar meson $\afn$ ($\tilde{O}_{\afn}=\lambdabar\lambda$), the pseudoscalar
meson $\aetap$ ($\tilde{O}_{\aetap}=\lambdabar\gamma_5\lambda$), the scalar
($0^{++}$) glueball, and the fermionic gluino-glue state $g\tilde{g}$
($\tilde{O}_{g\tilde{g}} = \sum_{\mu\nu} \sigma_{\mu\nu} \textrm{Tr} \left[
F^{\mu\nu} \lambda\right]$), see~\cite{Ali:2018dnd}. The scalar glueball and
the $\afn$ meson are combined in a common variational basis for the scalar
channel. The lightest states are expected to form a chiral supermultiplet,
which consists of a scalar, a pseudoscalar, and a fermionic spin 1/2
particle. From our previous investigations we expect a reasonable overlap of
both the $\afn$ and the scalar glueball with the lightest scalar state,
whereas the lightest pseudoscalar state seems to have a dominant overlap
with the $\aetap$ rather than with the $0^{-+}$ glueball. Therefore we
consider the meson-glueball mixing only in the scalar channel, and neglect,
at the moment, the $0^{-+}$ glueball. Note that the measurement of the
particle masses in SYM is quite challenging, involving only flavour singlet and glueball states. 

The chiral extrapolations to the point of vanishing adjoint pion mass $m_{\api}$ are shown in Fig.~\ref{fig:chiralcoarse}.
Away from the chiral point, the particles have different masses and the
chiral multiplet splits. This splitting is sizable at least for the coarsest lattice spacings. 
At these coarsest lattices, the gluino-glue becomes the heaviest particle, whereas the scalar particle becomes the lightest state.
There is an indication of a remaining mass splitting in the chiral
limit at the three coarsest lattice spacings.

At our two finest lattice spacings ($\beta=5.5$ and
$\beta=5.6$), there is no considerable splitting between the states of the
multiplet in the chiral limit. 
The scalar, pseudoscalar, and fermion masses are degenerate within errors at $\beta=5.6$ \footnote{We have neglected the ensemble with the largest adjoint pion mass at $\beta=5.6$ ($\kappa=0.1645$) for the extrapolations of the spectrum since the masses of the gluino-glue and the $\aetap$ deviated too strongly from the assumed linear behavior in $m_{\api}^2$. 
This manifests itself in a large $\chi^2_r\approx7$ and $\chi^2_r\approx2.8$ for the gluino-glue and $\aetap$, respectively, when fitting their masses at $\beta=5.6$ in lattice units linearly to the chiral limit.  It is the ensemble with the largest physical adjoint pion mass ($(\wchiral m_{\api})^2\approx3.1$) and therefore it is plausible that higher terms in $m_{\api}^2$ become relevant.}. The $0^{++}$ state
has the largest error of around $20\%$, and it can not be expected to be more precise
than the current glueball measurements in QCD.

A particular problem with our first data at the finest lattice spacing
($\beta=5.6$) has been the long autocorrelation due to topological freezing.
As we have already shown in our previous publication \cite{Ali:2018dnd}, larger volumes allow for more topological fluctuations, but the autocorrelation time of quantities like $w_0$ is still considerably large.

The three different lattice spacings allow for the first time a complete
extrapolation of the lightest states of SU(3) SYM to the continuum. Compared to our
previous work with an unimproved Wilson fermion action for the
investigations of SU(2) SYM, the differences of the masses in units of $\wchiral$
between the different lattice spacings are smaller and the continuum
extrapolation is rather flat thanks to the clover improved fermion action. Due
to the weak dependence on the lattice spacing, the linear and quadratic extrapolations are
consistent with each other, see Fig.~\ref{fig:continuumextr}. The final results using the two different fit procedures 
are summarised in the following table:
\begin{center}
\begin{tabular}{llll}
	\toprule
	 Fit  &     $w_0m_{g\tilde{g}}$ &   $w_0 m_{0^{++}}$ & $w_0 m_{\mathrm{a-}\eta'}$ \\
	\midrule
	linear fit  &   0.917(91) &  1.15(30) &  1.05(10) \\
 	quadratic fit &  0.991(55) &  0.97(18) &  0.950(63) \\
 	SU(2) SYM in~\cite{Ali:2019gzj}   & 0.93(6) & 1.3(2) & 0.98(6)  \\
	\bottomrule
\end{tabular}
\end{center}
For comparison, we have also added the data from our previous investigations of SU(2) SYM to the table.

We have finalised our first continuum extrapolation of the lightest bound
states in supersymmetric SU(3) Yang-Mills theory. We have found a formation of a chiral supermultiplet in the
continuum limit. In combination with the results from an analysis of the
supersymmetric Ward identities, this is a good indication for the absence of
supersymmetry breaking by the non-perturbative dynamics of the theory. It
also shows that the unavoidable breaking of supersymmetry by the lattice
discretisation is under control in this non-trivial theory.

This important observation opens the way towards several 
further investigations of SU(3) SYM, in particular concerning the phase transitions
and chiral dynamics of the theory. In addition, it is the first step towards 
investigations of supersymmetric QCD and other supersymmetric gauge theories 
that can not be accomplished without control of the supersymmetry breaking in the 
pure gauge sector.

Our investigation is based on the approach proposed in \cite{Curci:1986sm}, which means that chiral
symmetry is broken in a Wilson discretisation.
Our data indicate that the symmetries are restored by a tuning of the gluino mass parameter 
and the approach can be considerably improved by the clover fermion action.

Our results can be compared to the our previous analysis of SU(2) SYM,
presented in~\cite{Bergner:2015adz,Ali:2019gzj}. We find that in units of $w_0$ the 
masses of the multiplets are compatible with each other. This indicates only
a weak dependence of the multiplet mass on $N_c$.

One interesting additional aspect for further investigations is the continuum limit of the splitting of the multiplet as a function of the soft supersymmetry breaking. 
Our current data in Fig.~\ref{fig:chiralcoarse} show that the slope of the bound state masses as a function of the gluino mass has 
a significant dependence on the lattice spacing. Therefore the continuum extrapolations away from the chiral limit are more challenging and we plan further investigations in this direction.

\section{Acknowledgements}
The authors gratefully acknowledge the Gauss Centre for Supercomputing
e.\,V.\,\linebreak(www.gauss-centre.eu) for funding this project by providing
computing time on the GCS Supercomputers JUQUEEN, JURECA, and JUWELS at J\"ulich Supercomputing
Centre (JSC) and SuperMUC at Leibniz Supercomputing Centre (LRZ). Further
computing time has been provided the compute cluster PALMA of the University of M\"unster. This work is
supported by the Deutsche Forschungsgemeinschaft (DFG) through the Research
Training Group ``GRK 2149: Strong and Weak Interactions - from Hadrons to
Dark Matter''. G.~Bergner acknowledges support from the Deutsche
Forschungsgemeinschaft (DFG) Grant No.\ BE 5942/2-1. S.~Ali acknowledges
financial support from the Deutsche Akademische Austauschdienst (DAAD).


\newpage{\pagestyle{empty}\cleardoublepage}  

\appendix 

\begin{widetext}
\section{Summary of the data}
\label{sec:data}
\begin{table*}[htb]
\begin{small}
	\begin{tabular}{llllllllll}
\toprule
$\beta$ &  $\kappa$ &  Volume &    $am_{\api}$ &       $w_0/a$ &      $r_0/a$ & $am_{g\tilde{g}}$ &   $am_{0^{++}}$ & $am_{\aetap}$ & \# configs\\
\midrule
 5.4 &  0.1692 &  $16^3\times32$  &0.6954(14)  &  1.1259(42) &  -         &  1.2164(79) &  0.443(37) &  0.771(16) & 7788 \\
 5.4 &  0.1695 &  $12^3\times24$  &0.6241(17)  &  1.1955(87) &  3.4(10)   &  1.090(17)  &  0.383(40) &  0.705(10) & 7628 \\
 5.4 &  0.1695 &  $16^3\times32$  &0.6298(24)  &  1.1819(38) &  -         &  1.089(20)  &  0.361(41) &  0.650(16) & 3980 \\
 5.4 &  0.1697 &  $16^3\times32$  &0.5759(14)  &  1.2711(53) &  -         &  1.003(13)  &  0.404(49) &  0.653(17) & 3892 \\
 5.4 &  0.1700 &  $12^3\times24$  &0.4933(22)  &  1.386(14)  &  4.30(27)  &  0.866(14)  &  0.372(37) &  0.553(16) & 5964 \\
 5.4 &  0.1700 &  $16^3\times32$  &0.4952(21)  &  1.3988(91) &  -         &  0.884(12)  &  0.370(61) &  0.594(16) & 5856 \\
 5.4 &  0.1703 &  $12^3\times24$  &0.3718(30)  &  1.693(15)  &   5.10(26) &  0.712(17)  &  0.367(46) &  0.459(18) & 4028 \\
 5.4 &  0.1703 &  $16^3\times32$  &0.3911(16)  &  1.6039(93) &  -         &  0.7523(78) &  0.369(35) &  0.5029(82)& 9820 \\
 5.4 &  0.1705 &  $12^3\times24$  &0.318(11)   &  1.767(30)  &  -         &  0.657(24)  &  -         &  0.439(19) & 3300 \\
 5.4 &  0.1705 &  $16^3\times32$  &0.3123(33)  &  1.7316(83) &  -         &  0.661(12)  &  0.404(38) &  0.428(18) & 8052 \\
 5.45 & 0.1685 &  $16^3\times32$  &0.4748(22)  &  1.6733(64) &  5.04(18)  &  0.7579(92) &  0.442(56) &  0.568(18) & 1312 \\
 5.45 & 0.1687 &  $16^3\times32$  &0.4232(18)  &  1.778(10)  &  5.48(13)  &  0.680(11)  &  0.458(55) &  0.523(16) & 2564 \\
 5.45 & 0.1690 &  $16^3\times32$  &0.3404(21)  &  2.036(11)  &  5.98(17)  &  0.5867(93) &  0.365(61) &  0.451(17) & 2480 \\
 5.45 & 0.1692 &  $16^3\times32$  &0.2699(31)  &  2.184(12)  &  6.82(31)  &  0.506(13)  &  0.409(34) &  0.390(18) & 1648 \\
 5.45 & 0.1693 &  $16^3\times32$  &0.2282(36)  &  2.306(26)  &  6.71(22)  &  0.4876(60) &  0.305(35) &  0.405(33) & 2280 \\
 5.5 &  0.1667 &  $16^3\times32$  &0.5468(13)  &  1.881(12)  &  5.56(14)  &  0.7573(79) &  0.559(71) &  0.6090(61)& 4116 \\
 5.5 &  0.1673 &  $16^3\times32$  &0.4344(13)  &  1.996(12)  &  6.15(16)  &  0.6387(89) &  0.416(47) &  0.485(17) & 3732 \\
 5.5 &  0.1678 &  $16^3\times32$  &0.3245(11)  &  2.247(10)  &  6.82(33)  &  0.507(12)  &  0.405(34) &  0.3996(99)& 7736 \\
 5.5 &  0.1680 &  $16^3\times32$  &0.2606(29)  &  2.470(17)  &  8.06(34)  &  0.440(12)  &  0.316(51) &  0.364(12) & 1888 \\
 5.6 &  0.1645 &  $24^3\times48$  &0.51380(42) &  2.3316(95) &  6.940(46) &  0.6164(42) &  0.479(69) &  0.5660(72)& 5952 \\
 5.6 &  0.1650 &  $24^3\times48$  &0.43677(70) &  2.548(15)  &  7.329(52) &  0.5581(82) &  0.500(62) &  0.504(12) & 5728 \\
 5.6 &  0.1655 &  $24^3\times48$  &0.34561(76) &  2.794(17)  &  8.005(58) &  0.4506(65) &  0.381(40) &  0.420(13) & 4688 \\
 5.6 &  0.1660 &  $24^3\times48$  &0.22819(86) &  3.141(25)  &  9.13(16)  &  0.3531(68) &  0.341(46) &  0.310(14) & 4096 \\

\bottomrule
\end{tabular}
\end{small}
\caption{This table contains a summary of the masses in lattice units: the gluino-glue mass, the mass of the $0^{++}$ state, which includes a mixing of the mesonic $\afn$ and the glueball, as well as the mass of the $\aetap$.
The mesonic $\aetap$ represents the mass in the $0^{-+}$ channel since the $0^{-+}$ glueball does not provide a relevant signal for the ground state mass. In addition the scales $w_0$ and $r_0$ are presented.
Since the extrapolations have been done with the $w_0$ scale, the $r_0$ values are not determined from the complete statistics. The last column contains the number of thermalised configurations.}
\end{table*}
\begin{table*}[htb]
\begin{small}

\begin{tabular}{lll}
\toprule
               $\beta$ &      $\wchiral/a$ &     $r_{0,\chi}/a$ \\
\midrule
 5.4 &  2.049(18) &  6.10(60) \\
 5.45 &  2.577(43) &  7.24(22)  \\
 5.5 &  2.860(39) &  7.93(29)  \\
 5.6 &  3.485(71) &  9.40(14)  \\
\bottomrule
\end{tabular}

\end{small}
\caption{This table summarises the chiral extrapolations of the scales $r_0$ and $w_0$, which are used in the chiral extrapolations of masses and to estimate the lattice spacing in physical units. 
The results in units of the Sommer parameter $r_0$
can be converted to QCD units, fm or MeV, using the QCD relation
$r_0=0.5$\,fm. The methods for the determination of $r_0$ from a fit of
the static quark-antiquark potential are explained in our earlier work on
SU(2) SYM~\cite{Bergner:2015adz}. 
The chiral values $\wchiral$ and $r_{0,\chi}$ of the
scales  are obtained at each $\beta$ by a fit of the data to a 
second order polynomial in $(a m_{\api})^2$.
}
\end{table*}
\end{widetext}
\end{document}